# Analytic Representation of The Square-Root Operator

Tepper L. Gill[1,2,3] and W. W. Zachary[1,2]

[1]Department of Electrical & Computer Engineering
[2]Computational Physics Laboratory
[3]Department of Mathematics
Howard University
Washington, D. C. 20059

**Abstract**

In this paper, we use the theory of fractional powers of linear operators to construct a general (analytic) representation theory for the square-root energy operator of relativistic quantum theory, which is valid for all values of the spin. We focus on the spin 1/2 case, considering a few simple yet solvable and physically interesting cases, in order to understand how to interpret the operator. Our general representation is uniquely determined by the Green's function for the corresponding Schrödinger equation. We find that, in general, the operator has a representation as a nonlocal composite of (at least) three singularities. In the standard interpretation, the particle component has two negative parts and one (hard core) positive part, while the antiparticle component has two positive parts and one (hard core) negative part. This effect is confined within a Compton wavelength such that, at the point of singularity, they cancel each other providing a finite result. Furthermore, the operator looks like the identity outside a few Compton wavelengths (cutoff). To our knowledge, this is the first example of a physically relevant operator with these properties.

When the magnetic field is constant, we obtain an additional singularity, which could be interpreted as particle absorption and emission. The physical picture that emerges is that, in addition to the confined singularities and the additional attractive (repulsive) term, the effective mass of the composite acquires an oscillatory behavior.

We also derive an alternate relationship between the Dirac equation (with minimal coupling) and the square-root equation that is much closer than the one obtained via the Foldy-Wouthuysen method, in that there is no change in the wave function. This is accomplished by considering the scalar potential to be a part of the mass. This approach leads to a new Klein-Gordon equation and a new square-root equation, both of which have the same eigenvalues and eigenfunctions as the Dirac equation. Finally, we develop a perturbation theory that allows us to extend the range of our theory to include suitable spacetime-dependent potentials.

## I. Background

**Introduction**

In the transition from nonrelativistic to relativistic quantum theory, the Hamiltonian

$$H = \frac{[\mathbf{p} - (e/c)\mathbf{A}]^2}{2m} + V$$

is replaced by the square-root equation:

$$H = \sqrt{c^2[\mathbf{p} - (e/c)\mathbf{A}]^2 + m^2 c^4} + V.$$

It is quite natural to expect that the first choice for a relativistic wave equation would be:

$$i\hbar \frac{\partial \psi}{\partial t} = \left[ \sqrt{c^2[\mathbf{p} - (e/c)\mathbf{A}]^2 + m^2 c^4} + V \right] \psi,$$

where $\mathbf{p} = -i\hbar \nabla$. However, no one knew how to directly relate this equation to physically important problems. Furthermore, this equation is nonlocal, meaning, in the terminology of the times (1920-30), that it is represented by a power series in the momentum operator. Historically, Schrödinger [1], Gordon [2], Klein [3] and others [4,5,6] attempted to circumvent this problem by starting with the relationship:

$$(H - V)^2 = c^2\left(\mathbf{p} - \frac{e}{c}\mathbf{A}\right)^2 + m^2 c^4,$$

which led to the Klein-Gordon equation. At that time, the hope was to construct a relativistic quantum theory that would provide a natural extension of the nonrelativistic case. However, the problems with the Klein-Gordon equation were so great that many investigators became frustrated and it was dropped from serious consideration for a few years.

Dirac [7] argued that the proper equation should be first order in both the space and time variables in order to be a true relativistic wave equation, and this led to the well-known Dirac equation.

**Purpose**

In a survey article on relativistic wave equations, Foldy [8] pointed out that, in the absence of interaction, the above equation "gives a perfectly good wave equation for the description of a (spin zero) free particle". Foldy [9] had shown in an earlier paper that the square-root form:

$$H = \boldsymbol{\beta} \sqrt{c^2 \mathbf{p}^2 + m^2 c^4},$$

provides a canonical representation for particles of all finite spin. However, when **A** is nonzero, the noncommutativity of **p** and **A** "appeared" to make it impossible to give an unambiguous meaning to the radical operator.

In this paper, we take a new look at this equation and the general problem of its relationship to the Klein-Gordon equation. First, we investigate the extent that the non-commutativity of **p** and **A** affect our ability to give an unambiguous meaning to the square-root operator. We show that a unique analytic representation is well defined for suitable time-independent **A** and $m$, provided we can solve a corresponding equation of the Schrödinger type. We then investigate a few simple cases of solvable models in order to get a feel for the physical interpretation of this operator. Finally, motivated by previous work[10], we show that the relationship of the square-root operator equation to the Klein-Gordon equation depends explicitly on the Minkowski postulate. Dropping this postulate, we derive another Klein-Gordon type equation that does not depend on this postulate and gives a direct relationship of the square-root equation to the Dirac equation (without the use of a unitary transformation).

To begin, we start with the equation:

$$S[\psi] = \mathbf{H}_s \psi = \left\{ \boldsymbol{\beta}\sqrt{c^2\left(\mathbf{p} - \tfrac{e}{c}\mathbf{A}\right)^2 - e\hbar c \boldsymbol{\Sigma} \cdot \mathbf{B} + m^2 c^4} \right\}\psi, \quad (1a)$$

where β and Σ are the Dirac matrices; $\boldsymbol{\beta} = \begin{bmatrix} \mathbf{I} & 0 \\ 0 & -\mathbf{I} \end{bmatrix}$, $\boldsymbol{\Sigma} = \begin{bmatrix} \boldsymbol{\sigma} & 0 \\ 0 & \boldsymbol{\sigma} \end{bmatrix}$; **I** and σ are the identity and Pauli matrices respectively. In Section II, we construct an analytic representation for (1a), under special conditions. After exploring simplifications in Sections III, VI and V, we argue that the special conditions are physically reasonable. In Section IIV, we construct the general solution to the equation:

$$i\hbar \frac{\partial}{\partial t}\psi = \left\{ \boldsymbol{\beta}\sqrt{c^2\left(\mathbf{p} - \tfrac{e}{c}\mathbf{A}\right)^2 - e\hbar c \boldsymbol{\Sigma} \cdot \mathbf{B} + m^2 c^4} \right\}\psi. \quad (1b)$$

In Section IIV, we show that if we treat the potential energy as a part of the mass, there is an alternate connection between the Dirac and square-root equations. In the conclusion, we summarize our results and discuss open problems. In the appendix, we summarize the basic theory of semigroups of operators and fractional powers of closed linear operators so that the paper is self-contained.

**II General Representation Ansatz**

In this section, we construct a general analytic representation for equation (1a). To make our approach clear, set $\mathbf{G} = -\left(\mathbf{p} - \tfrac{e}{c}\mathbf{A}\right)^2$ and $\omega^2 = m^2 c^2 - \tfrac{e\hbar}{c}\boldsymbol{\Sigma} \cdot \mathbf{B}$. Except for time independence, we leave the form of the vector potential and the mass unspecified. We assume that $-\mathbf{G} + \omega^2$ satisfies the conditions required to be a generator of a unitary group (selfadjoint). Using the above notation, we can write (1a) as

$$S[\psi] = \left\{ c\boldsymbol{\beta}\sqrt{-\mathbf{G} + \omega^2} \right\}\psi. \tag{2}$$

Using the analytic theory of fractional powers of closed linear operators (see the Appendix, equation (A8)), it can be shown that, for generators of unitary groups, we can write (2) as: (using $\sqrt{\mathbf{F}} = (1/\sqrt{\mathbf{F}})\mathbf{F}$)

$$S[\psi] = \frac{c\boldsymbol{\beta}}{\pi}\int_0^\infty \left[(\lambda + \omega^2) - \mathbf{G}\right]^{-1}(-\mathbf{G} + \omega^2)\psi \frac{d\lambda}{\sqrt{\lambda}}, \tag{3}$$

where $\left[(\lambda + \omega^2) - \mathbf{G}\right]^{-1}$ is the resolvent associated with the operator $(\mathbf{G} - \omega^2)$. The resolvent can be computed directly if we can find the fundamental solution to the equation

$$\partial Q(\mathbf{x},\mathbf{y};t)/\partial t + (\mathbf{G} - \omega^2)Q(\mathbf{x},\mathbf{y};t) = \delta(\mathbf{x} - \mathbf{y}) \tag{4}$$

It is shown in Schulman[11] that the equation:

$$i\hbar \partial \bar{Q}(\mathbf{x},\mathbf{y};t)/\partial t + (\tfrac{1}{2M}\mathbf{G} - V)\bar{Q}(\mathbf{x},\mathbf{y};t) = \delta(\mathbf{x} - \mathbf{y}) \tag{5}$$

has the general (infinitesimal) solution

$$\bar{Q}(\mathbf{x},\mathbf{y};t) = \left(\frac{M}{2\pi i\hbar t}\right)^{3/2} \exp\left\{\frac{it}{\hbar}\left[\frac{M}{2}\left(\frac{\mathbf{x}-\mathbf{y}}{t}\right)^2 - V(\mathbf{y})\right] + \frac{ie}{\hbar c}(\mathbf{x}-\mathbf{y})\cdot \mathbf{A}[\tfrac{1}{2}(\mathbf{x}+\mathbf{y})]\right\}, \tag{6}$$

provided that $\mathbf{A}$ and $m$ are time-independent. Note that we used the midpoint evaluation in the last term of equation (6) ($\mathbf{A}[\tfrac{1}{2}(\mathbf{x}+\mathbf{y})]$). If we set $\omega^2/i\hbar = V$ and $M = i\hbar/2$, we get that

$$Q(\mathbf{x},t;\mathbf{y},0) = \int_{\mathbf{x}(0)=\mathbf{y}}^{\mathbf{x}(t)=\mathbf{x}} \mathcal{D}W_{\mathbf{x},t}[\mathbf{x}(s)] \exp\left\{\int_0^t V[\mathbf{x}(s)]ds + \frac{ie}{\hbar c}\int_{\mathbf{y}}^{\mathbf{x}} \mathbf{A}[\mathbf{x}(s)]\cdot d\mathbf{x}(s)\right\} \tag{7}$$

solves (4), where

$$\int_{\mathbf{x}(0)=\mathbf{y}}^{\mathbf{x}(t)=\mathbf{x}} \mathcal{D}W_{\mathbf{x},t}[\mathbf{x}(s)] = \int_{\mathbf{x}(0)=\mathbf{y}}^{\mathbf{x}(t)=\mathbf{x}} \mathcal{D}[\mathbf{x}(s)]\exp\left\{-\tfrac{1}{4}\int_0^t \left|\frac{d\mathbf{x}(s)}{ds}\right|^2 ds\right\}$$

$$= \lim_{N\to\infty}\left[\frac{1}{4\pi\varepsilon(N)}\right]^{nN/2} \int_{\mathbf{R}^n}\prod_{k=1}^N dx_j \exp\left\{-\sum_{j=1}^N \left[\frac{1}{4\varepsilon(N)}(x_j - x_{j-1})^2\right]\right\},$$

and $\varepsilon(N) = t/N$.

We now assume that we can write $\int_{\mathbf{y}}^{\mathbf{x}} \mathbf{A}[\mathbf{x}(s)] \cdot d\mathbf{x}(s) = \bar{\mathbf{A}} \cdot (\mathbf{x} - \mathbf{y})$ and $\int_{0}^{t} V[\mathbf{x}(s)] ds = \bar{V} t$ ($\bar{V}t = \omega^2 t/\hbar^2$) over the region of interest. (We will discuss the physical meaning of this assumption later.) Under these conditions, using (7), we can compute $\left[(\lambda + \omega^2) - \mathbf{G}\right]^{-1}$ from

$$\left[(\lambda + \omega^2) - \mathbf{G}\right]^{-1} f(\mathbf{x}) = \int_0^\infty e^{-\lambda t} \left[\int_{\mathbf{R}^3} Q(\mathbf{x}, t; \mathbf{y}, 0) f(\mathbf{y}) d\mathbf{y}\right] dt. \tag{8}$$

If we interchange the order of integration in (8) and use (7), we get

$$\left[(\lambda + \omega^2) - \mathbf{G}\right]^{-1} f(\mathbf{x})$$
$$= \int_{\mathbf{R}^3} \exp\left\{\tfrac{ie}{\hbar c}\bar{\mathbf{A}} \cdot (\mathbf{x} - \mathbf{y})\right\} \left\{\int_0^\infty \exp\left[-\frac{(\mathbf{x} - \mathbf{y})^2}{4t} - \frac{\omega^2 t}{\hbar^2} - \lambda t\right] \frac{dt}{(4\pi t)^{3/2}}\right\} f(\mathbf{y}) d\mathbf{y}. \tag{9a}$$

Using a table of Laplace transforms[12], the inner integral can be computed to get

$$\int_0^\infty \exp\left[-\frac{(\mathbf{x} - \mathbf{y})^2}{4t} - \frac{\omega^2 t}{\hbar^2} - \lambda t\right] \frac{dt}{(4\pi t)^{3/2}} = \frac{1}{4\pi} \frac{\exp\left[-\sqrt{(\lambda + \mu^2)} \|\mathbf{x} - \mathbf{y}\|\right]}{\|\mathbf{x} - \mathbf{y}\|}, \tag{9b}$$

where $\mu^2 = (\omega^2/\hbar^2)$. Equation (3) now becomes
$S[\psi](\mathbf{x})$

$$= \tfrac{c\beta}{4\pi^2} \int_0^\infty \left\{\int_{\mathbf{R}^3} \exp\left\{\tfrac{ie}{\hbar c}\bar{\mathbf{A}} \cdot (\mathbf{x} - \mathbf{y})\right\} \frac{\exp\left[-\sqrt{(\lambda + \mu^2)}\|\mathbf{x} - \mathbf{y}\|\right]}{\|\mathbf{x} - \mathbf{y}\|} (-\mathbf{G} + \omega^2) \psi(\mathbf{y}) d\mathbf{y}\right\} \tfrac{d\lambda}{\sqrt{\lambda}}. \tag{10}$$

Once again, we interchange the order of integration in (10) and perform the computations to get ($K_1[z]$ is the modified Bessel function of the third kind and first order)

$$\int_0^\infty \left\{\frac{\exp\left[-\sqrt{(\lambda + \mu^2)}\|\mathbf{x} - \mathbf{y}\|\right]}{\|\mathbf{x} - \mathbf{y}\|}\right\} \frac{d\lambda}{\sqrt{\lambda}} = \frac{4\mu \Gamma(\tfrac{3}{2})}{\pi^{1/2}} \frac{K_1[\mu\|\mathbf{x} - \mathbf{y}\|]}{\|\mathbf{x} - \mathbf{y}\|}. \tag{11}$$

Thus, if we set, $\mathbf{a} = \tfrac{e}{\hbar c} \mathbf{A}$ and $\bar{\mathbf{a}} = \tfrac{e}{\hbar c} \bar{\mathbf{A}}$ we get

$$S[\psi](\mathbf{x}) = \frac{c\beta}{2\pi^2} \int_{\mathbf{R}^3} \exp\left[i\bar{\mathbf{a}} \cdot (\mathbf{x} - \mathbf{y})\right] \frac{\mu K_1[\mu\|\mathbf{x} - \mathbf{y}\|]}{\|\mathbf{x} - \mathbf{y}\|} (-\mathbf{G} + \omega^2) \psi(\mathbf{y}) d\mathbf{y}. \tag{12}$$

Now,
$-\mathbf{G} + \omega^2 = \hbar^2 \left(-\Delta + 2i\mathbf{a} \cdot \nabla + i\nabla \cdot \mathbf{a} + \mathbf{a}^2 + \mu^2\right)$ so that (12) becomes

$$S[\psi](\mathbf{x}) = \tfrac{\hbar^2 c\beta}{2\pi^2} \int_{\mathbf{R}^3} e^{i\bar{\mathbf{a}}\cdot(\mathbf{x}-\mathbf{y})} \frac{\mu K_1[\mu\|\mathbf{x}-\mathbf{y}\|]}{\|\mathbf{x}-\mathbf{y}\|} \left(-\Delta + 2i\mathbf{a}\cdot\nabla + i\nabla\cdot\mathbf{a} + \mathbf{a}^2 + \mu^2\right)\psi(\mathbf{y})d\mathbf{y}. \qquad (13)$$

Before continuing, we recall a few standard results that will be used to compute (13). We assume that $f, g \in \mathbf{H}_0^2(\mathbf{D})$, $\mathbf{D} \subseteq \mathbf{R}^3$, $\partial\mathbf{D}$ smooth. From the divergence theorem we have that ( $\mathbf{v}$ is the outward normal on $\partial\mathbf{D}$)

$$\int_D fg_{y_i} d\mathbf{y} = \int_{\partial D}(fg)v_i d\mathbf{S} - \int_D gf_{y_i} d\mathbf{y}, \qquad (14)$$

$$\int_D f(\mathbf{a}\cdot\nabla)g d\mathbf{y} = \int_{\partial D}(fg)(\mathbf{a}\cdot\mathbf{v})d\mathbf{S} - \int_D g(\mathbf{a}\cdot\nabla)f d\mathbf{y}. \qquad (15)$$

We will also need Green's identity in the form

$$\int_D f\Delta g d\mathbf{y} = \int_{\partial D} fg_\nu d\mathbf{S} - \int_{\partial D} gf_\nu d\mathbf{S} + \int_D g\Delta f d\mathbf{y}.$$

Let us now consider a ball $\mathbf{B}_\rho(\mathbf{x})$ of radius $\rho$ about $\mathbf{x}$ so that $\mathbf{R}^3 = \mathbf{R}_\rho^3 \cup \mathbf{B}_\rho(\mathbf{x})$, where $\mathbf{R}_\rho^3 = \left(\mathbf{R}^3 \setminus \mathbf{B}_\rho(\mathbf{x})\right)$, so that $\partial\mathbf{R}_\rho^3 = \left(\partial\mathbf{R}^3 \setminus \partial\mathbf{B}_\rho(\mathbf{x})\right)$. Let $\mathbf{v}$ be the outward normal on $\partial\mathbf{R}_\rho^3$. It follows that $-\mathbf{v}$ is the outward normal on $\mathbf{B}_\rho(\mathbf{x})$ and $\mathbf{y} = \mathbf{x} - \mathbf{v}\rho$ on $\partial\mathbf{B}_\rho(\mathbf{x})$. Using (14) and (15), we can write (13) as ($u = \mu\|\mathbf{x}-\mathbf{y}\|$ and $\mathbf{T}[\psi](\mathbf{x}) = \tfrac{2\pi^2\beta}{\hbar^2 c} S[\psi](\mathbf{x})$)

$$\begin{aligned}
\mathbf{T}[\psi]_\rho &= \int_{\mathbf{R}_\rho^3} e^{i\bar{\mathbf{a}}\cdot(\mathbf{x}-\mathbf{y})}\mu^2\left[\mathbf{K}_1(u)/u\right]\left(-\Delta + 2i\mathbf{a}\cdot\nabla + i\nabla\cdot\mathbf{a} + \mathbf{a}^2 + \mu^2\right)\psi d\mathbf{y} \\
&= \int_{\mathbf{R}_\rho^3} e^{i\bar{\mathbf{a}}\cdot(\mathbf{x}-\mathbf{y})}\mu^2\left[\mathbf{K}_1(u)/u\right]\left(\mu^2 + i\nabla\cdot\mathbf{a} + \mathbf{a}^2\right)\psi d\mathbf{y} \\
&\quad -2i\int_{\mathbf{R}_\rho^3}\mathbf{a}\cdot\nabla\left\{e^{i\bar{\mathbf{a}}\cdot(\mathbf{x}-\mathbf{y})}\mu^2\left[\mathbf{K}_1(u)/u\right]\right\}\psi d\mathbf{y} - \int_{\mathbf{R}_\rho^3}\Delta\left\{e^{i\bar{\mathbf{a}}\cdot(\mathbf{x}-\mathbf{y})}\mu^2\left[\mathbf{K}_1(u)/u\right]\right\}\psi d\mathbf{y} \\
&\quad -\int_{\partial\mathbf{R}_\rho^3}\left\{e^{i\bar{\mathbf{a}}\cdot(\mathbf{x}-\mathbf{y})}\mu^2\left[\mathbf{K}_1(u)/u\right]\right\}_{-\nu}\psi d\mathbf{S} + \int_{\partial\mathbf{R}_\rho^3}\left\{e^{i\bar{\mathbf{a}}\cdot(\mathbf{x}-\mathbf{y})}\mu^2\left[\mathbf{K}_1(u)/u\right]\right\}\psi_{-\nu} d\mathbf{S} \\
&\quad +2i\int_{\partial\mathbf{R}_\rho^3} e^{i\bar{\mathbf{a}}\cdot(\mathbf{x}-\mathbf{y})}\mu^2\left[\mathbf{K}_1(u)/u\right]\psi(\mathbf{a}\cdot\mathbf{v})d\mathbf{S}.
\end{aligned} \qquad (16)$$

It is clear that the surface integrals vanish on $\partial\mathbf{R}^3$, so we need only consider them on $\partial\mathbf{B}_\rho(\mathbf{x})$. It is easy to check that the integrands in the last two terms are continuous so they vanish as $\rho \to 0$. Easy analysis shows that the only possible nonvanishing part of the remaining surface term is ($d\mathbf{S} = \rho^2 \sin\theta d\theta d\phi = u^2/\mu^2 d\Omega$)

$$\int_{\partial\mathbf{B}_\rho} \exp[i\bar{\mathbf{a}}\cdot\mathbf{v}\rho]\left[u^2/\mu^2\right]\left[\mathbf{K}_1(u)/u\right]_{-\nu}\psi d\Omega, \qquad (17)$$

where, on $\partial\mathbf{B}_\rho(\mathbf{x})$, $\|\mathbf{x}-\mathbf{y}\| = \rho$, $(x_i - y_i) = v_i$ & $u = \mu\rho$. We also have

$$\left[\frac{\mathbf{K}_1(u)}{u}\right]_{-\mathbf{v}} = -\mathbf{v}\cdot\nabla\left[\frac{\mathbf{K}_1(u)}{u}\right] = -\sum_{i=1}^{3} v_i \frac{d}{du}\left[\frac{\mathbf{K}_1(u)}{u}\right]\frac{\partial u}{\partial y_i},$$

and

$$\frac{d}{du}\left[\frac{\mathbf{K}_1(u)}{u}\right] = -\frac{\mathbf{K}_2(u)}{u}, \quad \& \quad \frac{\partial u}{\partial y_i} = \frac{\partial \mu}{\partial y_i}\|\mathbf{x}-\mathbf{y}\| - \mu\frac{(x_i - y_i)}{\|\mathbf{x}-\mathbf{y}\|},$$

so that

$$\left[\frac{\mathbf{K}_1(u)}{u}\right]_{-\mathbf{v}} = \frac{\mathbf{K}_2(u)}{u}[\rho\mathbf{v}\cdot\nabla\mu - \mu]. \tag{18}$$

Assuming that $\mu \in \mathbf{H}_0^2(\mathbf{R}^3)$, it is easy to see that $u^2[\mathbf{K}_2(u)/u]\mathbf{v}\cdot\nabla\mu\rho$ is continuous as $\rho \to 0$, so that the surface integral of this term vanishes. Thus we only need consider

$$\lim_{\rho\to 0}\int_{\|\mathbf{x}-\mathbf{y}\|=\rho}\exp\{i\bar{\mathbf{a}}[\mathbf{x}-\mathbf{v}\rho/2]\cdot\mathbf{v}\rho\}[u/\mu(\mathbf{x}-\mathbf{v}\rho)]\psi(\mathbf{x}-\mathbf{v}\rho)\mathbf{K}_2(u)d\Omega.. \tag{19}$$

The first three terms in the above integral are continuous as $\rho \to 0$. However, as $u = \mu\rho$ and $\mathbf{K}_2(u) \approx 1/u^2$ near zero, we see that because of the last term, the integral diverges like $1/u$. This is the first of several divergent integrals that arise in the analytic representation of the square root operator. For later use, we represent it as

$$4\pi\int_{\mathbf{R}^3}e^{i\bar{\mathbf{a}}\cdot(\mathbf{x}-\mathbf{y})}\mu\frac{\mathbf{K}_2[\mu\|\mathbf{x}-\mathbf{y}\|]}{\mu\|\mathbf{x}-\mathbf{y}\|}\delta(\mathbf{x}-\mathbf{y})\psi(\mathbf{y})d\mathbf{y}. \tag{20}$$

In the limit as $\rho \to 0$, equation (16) becomes:

$$\mathbf{T}[\psi] = \int_{\mathbf{R}^3}\left(\mu^2 + i\nabla\cdot\mathbf{a} + \mathbf{a}^2\right)e^{i\bar{\mathbf{a}}\cdot(\mathbf{x}-\mathbf{y})}\mu^2\frac{\mathbf{K}_1[\mu\|\mathbf{x}-\mathbf{y}\|]}{\mu\|\mathbf{x}-\mathbf{y}\|}\psi d\mathbf{y}$$

$$-2i\int_{\mathbf{R}^3}\mathbf{a}\cdot\nabla\left\{e^{i\bar{\mathbf{a}}\cdot(\mathbf{x}-\mathbf{y})}\mu^2\frac{\mathbf{K}_1[\mu\|\mathbf{x}-\mathbf{y}\|]}{\mu\|\mathbf{x}-\mathbf{y}\|}\right\}\psi d\mathbf{y} - \int_{\mathbf{R}^3}\Delta\left\{e^{i\bar{\mathbf{a}}\cdot(\mathbf{x}-\mathbf{y})}\mu^2\frac{\mathbf{K}_1[\mu\|\mathbf{x}-\mathbf{y}\|]}{\mu\|\mathbf{x}-\mathbf{y}\|}\right\}\psi d\mathbf{y} \tag{21}$$

$$+\int_{\mathbf{R}^3}e^{i\bar{\mathbf{a}}\cdot(\mathbf{x}-\mathbf{y})}\mu\frac{\mathbf{K}_2[\mu\|\mathbf{x}-\mathbf{y}\|]}{\mu\|\mathbf{x}-\mathbf{y}\|}4\pi\delta(\mathbf{x}-\mathbf{y})\psi d\mathbf{y}.$$

The above expression can be further refined after computation of the middle two terms. The calculations are long but straightforward, so we provide intermediate steps omitting details. Using $\Delta(fg) = f\Delta g + f\Delta g + 2\nabla f\cdot\nabla g$, the third term becomes

$$\int_{\mathbf{R}^3} \Delta \left\{ e^{i\overline{\mathbf{a}} \cdot (\mathbf{x}-\mathbf{y})} \mu \frac{\mathbf{K}_1[\mu \|\mathbf{x}-\mathbf{y}\|]}{\|\mathbf{x}-\mathbf{y}\|} \right\} \psi d\mathbf{y}$$

$$= \int_{\mathbf{R}^3} \Delta \{ e^{i\overline{\mathbf{a}} \cdot (\mathbf{x}-\mathbf{y})} \mu^2 \} \frac{\mathbf{K}_1[\mu \|\mathbf{x}-\mathbf{y}\|]}{\mu \|\mathbf{x}-\mathbf{y}\|} \psi d\mathbf{y} + \int_{\mathbf{R}^3} e^{i\overline{\mathbf{a}} \cdot (\mathbf{x}-\mathbf{y})} \mu^2 \Delta \left\{ \frac{\mathbf{K}_1[\mu \|\mathbf{x}-\mathbf{y}\|]}{\mu \|\mathbf{x}-\mathbf{y}\|} \right\} \psi d\mathbf{y} \quad (22)$$

$$+ 2 \int_{\mathbf{R}^3} \nabla \{ e^{i\overline{\mathbf{a}} \cdot (\mathbf{x}-\mathbf{y})} \mu^2 \} \cdot \nabla \left\{ \frac{\mathbf{K}_1[\mu \|\mathbf{x}-\mathbf{y}\|]}{\mu \|\mathbf{x}-\mathbf{y}\|} \right\} \psi d\mathbf{y}.$$

For further refinement, we need the following: ($u = \mu \|\mathbf{x}-\mathbf{y}\|$, $w = e^{i\overline{\mathbf{a}} \cdot (\mathbf{x}-\mathbf{y})} \mu^2$)

$$\Delta[\mathbf{K}_1(u)/u] = [\mathbf{K}_3(u)/u](\nabla u)^2 - [\mathbf{K}_2(u)/u](\Delta u),$$

$$\nabla[\mathbf{K}_1(u)/u] = -\nabla u [\mathbf{K}_2(u)/u],$$

$$\nabla u = (\|\mathbf{x}-\mathbf{y}\| \nabla \mu - \mu[(\mathbf{x}-\mathbf{y})/\|\mathbf{x}-\mathbf{y}\|]) = u\left(\frac{\nabla \mu}{\mu}\right) - \mu^2 \frac{(\mathbf{x}-\mathbf{y})}{u}, \quad (23)$$

$$(\nabla u)^2 = \mu^2 + \|\mathbf{x}-\mathbf{y}\|^2 (\nabla \mu)^2 - 2\mu[\nabla \mu \cdot (\mathbf{x}-\mathbf{y})],$$

$$\Delta u = \|\mathbf{x}-\mathbf{y}\| \Delta \mu - 2[\nabla \mu \cdot (\mathbf{x}-\mathbf{y}) - \mu]/\|\mathbf{x}-\mathbf{y}\|.$$

$$\begin{aligned} K_3[u]/u &= K_1[u]/u + 4 K_2[u]/u^2 \\ K_2[u]/u &= K_0[u]/u + 2 K_1[u]/u^2 \end{aligned} \quad (24)$$

Using (23), we can compute (22) and the second term of (21) to get ($\mathbf{z} = \mathbf{x} - \mathbf{y}$)

$$\int_{\mathbf{R}^3} \Delta \left\{ w \frac{K_1[u]}{u} \right\} \psi d\mathbf{y} = \int_{\mathbf{R}^3} \Delta w \frac{K_1[u]}{u} \psi d\mathbf{y} + \int_{\mathbf{R}^3} w \left[ \mu^2 + u^2 \frac{(\nabla \mu)^2}{\mu^2} - 2[\nabla \mu \cdot \mathbf{u}] \right] \frac{K_3[u]}{u} \psi d\mathbf{y}$$
$$- \int_{\mathbf{R}^3} w \left[ u \frac{\Delta \mu}{\mu} - 2 \frac{[\nabla \mu \cdot \mathbf{u} - \mu^2]}{u} \right] \frac{K_2[u]}{u} \psi - 2 \int_{\mathbf{R}^3} \left( u \left( \frac{\nabla w \cdot \nabla \mu}{\mu} \right) - \mu \frac{\nabla w \cdot \mathbf{u}}{u} \right) \frac{K_2[u]}{u} \psi d\mathbf{y}, \quad (25)$$

$$\int_{\mathbf{R}^3} \mathbf{a} \cdot \nabla \left\{ w \frac{K_1[u]}{u} \right\} \psi d\mathbf{y} = + \int_{\mathbf{R}^3} (\mathbf{a} \cdot \nabla w) \frac{K_1[u]}{u} \psi d\mathbf{y} - \int_{\mathbf{R}^3} w \left\{ u \frac{(\mathbf{a} \cdot \nabla \mu)}{\mu} - \mu \frac{\mathbf{a} \cdot \mathbf{u}}{u} \right\} \frac{K_2[u]}{u} \psi d\mathbf{y}. \quad (26)$$

In order to complete our representation for the square root operator, we need to compute $\nabla w$ & $\Delta w$:

$$\nabla w = w \left\{ 2 \frac{\nabla \mu}{\mu} + i[\nabla(\overline{\mathbf{a}} \cdot \mathbf{z})] \right\} = w \left\{ 2 \frac{\nabla \mu}{\mu} + i[(\mathbf{z} \cdot \nabla) \cdot \overline{\mathbf{a}}] - i\overline{\mathbf{a}} \right\},$$

$$\Delta w = w \left\{ 2 \left[ \frac{\Delta \mu}{\mu} + \left( \frac{\nabla \mu}{\mu} \right)^2 \right] + 4i \left[ \frac{1}{\mu} \nabla(\overline{\mathbf{a}} \cdot \mathbf{z}) \cdot \nabla \mu \right] + i[\Delta(\overline{\mathbf{a}} \cdot \mathbf{z})] - [\nabla(\overline{\mathbf{a}} \cdot \mathbf{z})]^2 \right\}. \quad (27)$$

It follows that

$$\mathbf{z} \cdot \nabla w = w\left\{2\tfrac{\mathbf{z}\cdot\nabla\mu}{\mu} + i\left[(\mathbf{z}\cdot\nabla)\overline{\mathbf{a}}\right]\cdot\mathbf{z} - i\mathbf{z}\cdot\overline{\mathbf{a}}\right\},$$
$$\left(\tfrac{\nabla w \cdot \nabla \mu}{\mu}\right) = w\left\{2\tfrac{(\nabla\mu)^2}{\mu^2} + i\left[(\mathbf{z}\cdot\nabla)\overline{\mathbf{a}}\right]\cdot\left(\tfrac{\nabla\mu}{\mu}\right) - i\overline{\mathbf{a}}\cdot\left(\tfrac{\nabla\mu}{\mu}\right)\right\} \tag{28}$$

and

$$S[\psi] = \frac{\hbar^2 \mu^2 c\beta}{2\pi^2}\left\{\int_{\mathbf{R}^3}(\mu^2 + i\nabla\cdot\mathbf{a} + \mathbf{a}^2)e^{i\overline{\mathbf{a}}\cdot\mathbf{z}}\frac{\mathbf{K}_1[\mu\|\mathbf{z}\|]}{\mu\|\mathbf{z}\|}\psi d\mathbf{y} + \int_{\mathbf{R}^3}e^{i\overline{\mathbf{a}}\cdot\mathbf{z}}\mu\frac{\mathbf{K}_2[\mu\|\mathbf{z}\|]}{\mu\|\mathbf{z}\|}4\pi\delta(\mathbf{z})\psi d\mathbf{y}\right.$$
$$-2i\int_{\mathbf{R}^3}(\overline{\mathbf{a}}\cdot\nabla w)\frac{\mathbf{K}_1[\mu\|\mathbf{z}\|]}{\mu\|\mathbf{z}\|}\psi d\mathbf{y} + 2i\int_{\mathbf{R}^3}w\left\{\|\mathbf{z}\|(\mathbf{a}\cdot\nabla\mu) - \mu\tfrac{\mathbf{a}\cdot\mathbf{z}}{\|\mathbf{z}\|}\right\}\frac{\mathbf{K}_2[\mu\|\mathbf{z}\|]}{\mu\|\mathbf{z}\|}\psi d\mathbf{y}$$
$$-\int_{\mathbf{R}^3}\Delta w\frac{\mathbf{K}_1[\mu\|\mathbf{z}\|]}{\mu\|\mathbf{z}\|}\psi d\mathbf{y} - \int_{\mathbf{R}^3}w\left[\mu^2 + \|\mathbf{z}\|^2(\nabla\mu)^2 - 2\mu[\nabla\mu\cdot\mathbf{z}]\right]\frac{\mathbf{K}_3[\mu\|\mathbf{z}\|]}{\mu\|\mathbf{z}\|}\psi d\mathbf{y}$$
$$+\int_{\mathbf{R}^3}w\left[\|\mathbf{z}\|\Delta\mu - 2\tfrac{[\nabla\mu\cdot\mathbf{z}-\mu]}{\|\mathbf{z}\|}\right]\frac{\mathbf{K}_2[\mu\|\mathbf{z}\|]}{\mu\|\mathbf{z}\|}\psi d\mathbf{y} + 2\int_{\mathbf{R}^3}\left(\|\mathbf{z}\|(\nabla w\cdot\nabla\mu) - \mu\tfrac{\nabla w\cdot\mathbf{z}}{\|\mathbf{z}\|}\right)\frac{\mathbf{K}_2[\mu\|\mathbf{z}\|]}{\mu\|\mathbf{z}\|}\psi d\mathbf{y}\left.\right\}. \tag{29}$$

Equation (29) allows us to explore the physical consequences for a number of possible combinations of (time-independent) vector potentials. However, we must first establish the conditions under which our initial assumptions can be expected to hold.

### III. The Free Particle Case ($\mu$ constant and $\mathbf{A} = 0$)

The free particle case is the simplest ($\mu$ constant and $\mathbf{A} = 0$). In this case, (29) reduces to

$$S[\psi](\mathbf{x}) = -\frac{2\mu^2\hbar^2 c\beta}{\pi^2}\int_{\mathbf{R}^3}\left[\frac{1}{\|\mathbf{x}-\mathbf{y}\|} - \pi\delta(\mathbf{x}-\mathbf{y})\right]\frac{\mathbf{K}_2[\mu\|\mathbf{x}-\mathbf{y}\|]}{\|\mathbf{x}-\mathbf{y}\|}\psi(\mathbf{y})d\mathbf{y}. \tag{30}$$

Using $K_2[u]/u = K_0[u]/u + 2K_1[u]/u^2$, we see that (30) has the representation

$$S[\psi](\mathbf{x}) = -\frac{2\mu^2\hbar^2 c\beta}{\pi^2}\int_{\mathbf{R}^3}\left[\frac{1}{\|\mathbf{x}-\mathbf{y}\|} - \pi\delta(\mathbf{x}-\mathbf{y})\right]\left\{\frac{\mathbf{K}_0[\mu\|\mathbf{x}-\mathbf{y}\|]}{\|\mathbf{x}-\mathbf{y}\|} + \frac{2\mathbf{K}_1[\mu\|\mathbf{x}-\mathbf{y}\|]}{\mu\|\mathbf{x}-\mathbf{y}\|^2}\right\}\psi(\mathbf{y})d\mathbf{y}. \tag{31}$$

Gill[10] first derived equation (31) using the method of fractional powers of closed linear operators. In order to identify a possible physical interpretation for (31), it will be helpful to review some properties of the Bessel functions $K_0[u]$, $K_{1/2}[u]/u^{1/2}$ and $K_1[u]/u$. If $\mathbf{x} \neq \mathbf{y}$, the effective kernel in (31) is

$$\frac{\mathbf{K}_0[\mu\|\mathbf{x}-\mathbf{y}\|]}{\|\mathbf{x}-\mathbf{y}\|^2} + \frac{2\mathbf{K}_1[\mu\|\mathbf{x}-\mathbf{y}\|]}{\mu\|\mathbf{x}-\mathbf{y}\|^3}. \tag{32}$$

(Note that the integral of $\|\mathbf{x}-\mathbf{y}\|^{-2}$ is finite over $\mathbf{R}^3$.) We follow Gradshteyn and Ryzhik[12]. For $0 < u \ll 1$, we have that:

$$\left. \begin{aligned} \mathbf{K}_1[u]/u &= c_1\left[1+\theta_1(u)\right]u^{-2} \\ \mathbf{K}_{1/2}[u]/u^{1/2} &= [\sqrt{\pi/2}]u^{-1} \\ \mathbf{K}_0[u] &= c_0\left[1+\theta_0(u)\right]\ln u^{-1} \end{aligned} \right\}, \tag{33a}$$

where $\theta_0(u) \downarrow 0$, $\theta_1(u) \downarrow 0$, $u \downarrow 0$. On the other hand, for $u \gg 1$, we have:

$$\left. \begin{aligned} \frac{\mathbf{K}_1[u]}{u} &= c_1\left[1+\theta_1'(u)\right]\frac{\exp\{-u\}}{u^{3/2}} \\ \mathbf{K}_{1/2}[u]/u^{1/2} &= [\sqrt{\pi/2}]\frac{\exp\{-u\}}{u} \\ \mathbf{K}_0[u] &= c_0\left[1+\theta_0'(u)\right]\frac{\exp\{-u\}}{u^{1/2}} \end{aligned} \right\}. \tag{33b}$$

In this case, the functions $\theta_0'(u)$, $\theta_1'(u)$ converge $\downarrow 0$ as $u \uparrow \infty$.

Recall that $g^2 \exp\{-u\}/u$ is the well-known Yukawa potential[13], conjectured in 1935 in order to account for the short range of the nuclear interaction that was expected to have massive exchange particles (where $g$ represents the "charge" of the exchange field). Yukawa assumed that the range of the exchange field was $1/\mu \cong 1.4$ fermi, which led to a mass value of about 170 times that of the electron. Anderson and Neddermeyer[14] discovered what was believed to be Yukawa's meson with a mass of 207 times that of an electron in 1935. However, this particle interacted so weakly with nuclei and had such a long lifetime, it was rejected as a participant in the nuclear interaction. Finally, in 1947 Lattes et al[15] identified the π-meson (pion) with all the expected properties.

Looking at equation (33a) in the strength of singularity sense, we see that when $0 < u \ll 1$,

$$\mathbf{K}_1[u]/u > \mathbf{K}_1[u]/u^{1/2} \gg \mathbf{K}_0[u]. \tag{34}$$

The $\mathbf{K}_0[u]$ term is the weakest possible singularity in that: $\lim_{u \to 0}\{u^\varepsilon \mathbf{K}_0[u]\} = 0$, $\varepsilon > 0$. (In fact, it is an integrable singularity.) On the other hand, the $\mathbf{K}_1[u]/u$ term has the strongest singularity (as it diverges like $1/u^2$), while the Yukawa term is halfway between them (diverges like $1/u$). From equation (33b) we see that for large $u$, inequality (34) is reversed so that:

$$\mathbf{K}_0[u] > \mathbf{K}_{1/2}[u]/u^{1/2} > \mathbf{K}_1[u]/u. \tag{35}$$

Although all three terms in (35) have an exponential cutoff, the $\mathbf{K}_0[u]$ term has the longest range (see (33b)). If we include the $-\mu^2$ term in equation (32), it becomes:

$$-\mu^2 \frac{\mathbf{K}_0[\mu\|\mathbf{x}-\mathbf{y}\|]}{\|\mathbf{x}-\mathbf{y}\|^2} - 2\mu \frac{\mathbf{K}_1[\mu\|\mathbf{x}-\mathbf{y}\|]}{\|\mathbf{x}-\mathbf{y}\|^3}. \tag{36}$$

Thus, $-\mu^2 \mathbf{K}_0[u]$ has an extra factor of $\mu$, compared to the $\mathbf{K}_1[u]/u$ term, giving a value of $5 \times 10^{13}\, cm^{-1}$, assuming that the mass is that of an electron. Hence, although this term is the weakest of all possible singularities near $\mathbf{x} = \mathbf{y}$, it is asymptotically stronger by a factor of at least $10^9$ in the asymptotic region. Thus, for $u \gg 1$, we can replace equation (35) by:

$$\mathbf{K}_0[u] \gg \mathbf{K}_{1/2}[u]/u^{1/2} > \mathbf{K}_1[u]/u. \tag{35b}$$

**Discussion**

Equation (31) is the first known example of a physically relevant operator with an analytic representation as a composite of three singularities. In the standard interpretation, the particle component has two negative and one (hard core) positive part, while the antiparticle component has two positive and one (hard core) negative part. <u>This effect is confined within a Compton wavelength</u> such that, and at the point of singularity, the terms cancel each other providing the action of a well-defined operator. <u>Furthermore, the operator looks (almost) like the identity outside a Compton wavelength</u>, but has a residual instantaneous connection with all the particles in the universe at each point in time (spatially nonlocal). This suggests that the square-root operator might represent the inside of an extended object. We will return to this discussion after we study a few other exact representations.

**IV. The Constant Case ( A and $\mu$ constant)**

Since a tractable solution for equation (2) is difficult to find, it is of some comfort that a number of exact simple solutions are possible. These provide a wealth of physical insight into the nature of this operator during interaction. When $\mathbf{A}$ and $\mu$ are constant, we get another solvable problem, which is still of physical interest. In this case, since $\nabla \cdot \mathbf{A} = 0$ and $\mathbf{B} = \nabla \times \mathbf{A} = 0$, equation (29) becomes: ($\bar{\mathbf{a}} = \mathbf{a}$)

$$S[\psi](\mathbf{x}) = -\frac{\mu^2 \hbar^2 c \beta}{\pi^2} \int_{\mathbf{R}^3} e^{i\mathbf{a}\cdot(\mathbf{x}-\mathbf{y})} \left[ \frac{1}{\|\mathbf{x}-\mathbf{y}\|} - \frac{2\pi\delta(\mathbf{x}-\mathbf{y})}{[1+i\mathbf{a}\cdot(\mathbf{x}-\mathbf{y})]} \right] \left\{ [1 + i\mathbf{a}\cdot(\mathbf{x}-\mathbf{y})] \frac{\mathbf{K}_2[\mu\|\mathbf{x}-\mathbf{y}\|]}{\|\mathbf{x}-\mathbf{y}\|} \right\} \psi(\mathbf{y}) d\mathbf{y}$$
$$- \frac{\mu \hbar^2 c \beta \mathbf{a}^2}{2\pi^2} \int_{\mathbf{R}^3} e^{i\mathbf{a}\cdot(\mathbf{x}-\mathbf{y})} \frac{\mathbf{K}_1[\mu\|\mathbf{x}-\mathbf{y}\|]}{\|\mathbf{x}-\mathbf{y}\|} \psi(\mathbf{y}) d\mathbf{y} \quad (37)$$

Thus, when $\mathbf{A}$ and $\mu = mc/\hbar$ are both constant, we get two extra terms and a multiplicative exponential factor. The first term is purely imaginary and it is not hard to see that it is nonsingular at $\mathbf{x} = \mathbf{y}$ (by power counting). In fact, the term approaches a constant (nonzero) value) at $\mathbf{x} = \mathbf{y}$, with a value depending only on the magnitude of $\mathbf{A}$. Physically, we interpret such a term as representing the absorption of energy from a beam (source).

## V. The Constant Field Case ($\mu, \mathbf{B}$ constant)

In this case, assuming a constant magnetic field $\mathbf{B}$ and constant mass, we get another exact solution. Since $\mathbf{A}(\mathbf{z}) = \frac{1}{2} \mathbf{z} \times \mathbf{B}$, computation of the last integral on the right-hand side of equation (7) yields $\int_{\mathbf{y}}^{\mathbf{x}} \left[ \frac{1}{2} \mathbf{x}(s) \times \mathbf{B} \right] \cdot d\mathbf{x}(s) = 0$. Since $\nabla \cdot \mathbf{A} = 0$, $\mathbf{a} = \mathbf{a}(\mathbf{y}) = \frac{e}{\hbar c} \mathbf{A}(\mathbf{y})$ and $\mathbf{a}(\mathbf{y}) \cdot \mathbf{y} = 0$, equation (29) reduces to:

$$S[\psi](\mathbf{x}) = -\frac{\mu^2 \hbar^2 c \beta}{\pi^2} \int_{\mathbf{R}^3} \left[ \frac{1}{\|\mathbf{x}-\mathbf{y}\|} - \frac{2\pi\delta(\mathbf{x}-\mathbf{y})}{[1+i\mathbf{a}\cdot\mathbf{x}]} \right] \left\{ [1 + i\mathbf{a}\cdot\mathbf{x}] \frac{\mathbf{K}_2[\mu\|\mathbf{x}-\mathbf{y}\|]}{\|\mathbf{x}-\mathbf{y}\|} \right\} \psi(\mathbf{y}) d\mathbf{y}$$
$$- \frac{\mu \hbar^2 c \beta \mathbf{a}^2}{2\pi^2} \int_{\mathbf{R}^3} \frac{\mathbf{K}_1[\mu\|\mathbf{x}-\mathbf{y}\|]}{\|\mathbf{x}-\mathbf{y}\|} \psi(\mathbf{y}) d\mathbf{y}. \quad (38)$$

As in the previous section, we get two extra terms, but without the multiplicative exponential factor. The first term is again purely imaginary and the second one is attractive and nonsingular. Now however, that first term is singular at $\mathbf{x} = \mathbf{y}$ like $1/\|\mathbf{x}-\mathbf{y}\|$. This happens because, unlike equation (37), where $\mathbf{a} \cdot (\mathbf{x} - \mathbf{y})$ cancels a $\|\mathbf{x} - \mathbf{y}\|$ factor in the denominator of the $K_2$ kernel, in equation (38) the $\mathbf{a} \cdot \mathbf{x}$ does not affect the divergent term in the kernel. However, in this case, it is possible that $\mathbf{a}$ would try to align itself so that $\mathbf{a} \cdot \mathbf{x} = 0$. Such a process would not be stable, and would depend on the available energy in the system.

In addition, the effective mass $\mu$ is constant but matrix-valued with complex components. In this case, $\mu^2 = m^2 c^2 / \hbar^2 - \frac{e}{\hbar c} \Sigma \cdot \mathbf{B}$, so with:

$$\Sigma = \begin{pmatrix} \sigma & 0 \\ 0 & \sigma \end{pmatrix}; \quad \sigma_1 = \begin{pmatrix} 0 & 1 \\ 1 & 0 \end{pmatrix}, \quad \sigma_2 = \begin{pmatrix} 0 & -i \\ i & 0 \end{pmatrix}, \quad \sigma_3 = \begin{pmatrix} 1 & 0 \\ 0 & -1 \end{pmatrix}, \qquad (39)$$

$$\mu^2 = \begin{bmatrix} (\frac{m^2 c^2}{\hbar^2} - \frac{e}{\hbar c} B_3) \mathbf{I}_2 & \frac{ie}{\hbar c}(B_2 - iB_1) \mathbf{I}_2 \\ \frac{-ie}{\hbar c}(B_2 - iB_1) \mathbf{I}_2 & (\frac{m^2 c^2}{\hbar^2} + \frac{e}{\hbar c} B_3) \mathbf{I}_2 \end{bmatrix}. \qquad (40)$$

We can rewrite $\mu$ as $\mu = U |\mu|$, where $U$ is a isometric operator (a restricted unitary operator in $\mathbb{C}^4$, the 4-dim. complex space), $|\mu| = [\mu^* \mu]^{1/2}$, where $\mu^*$ is the Hermitian conjugate of $\mu$ and the square root is computed using elementary spectral theory.

From properties of Bessel functions we know that, for nonintegral $v$, we can represent $\mathbf{K}_v[u]$ as

$$(2/\pi) \mathbf{K}_v[u] = \frac{\mathbf{I}_{-v}(u) - \mathbf{I}_v(u)}{\sin \pi v} = \frac{e^{i/2(\pi v)} \mathbf{J}_{-v}(iu) - e^{-i/2(\pi v)} \mathbf{J}_v(iu)}{\sin \pi v}. \qquad (41)$$

In the limit as $v$ approaches an integer, equation (41) takes the indeterminate form $0/0$, and is defined via L'Hôpital's rule. However, for our purposes, we assume that $v$ is close to an integer and $u = u_1 + iu_2$, $u_2 \neq 0$. It follows that $\mathbf{K}_v[u]$ acquires some of the oscillatory behavior of $\mathbf{J}_v[z]$. Thus, we can interpret equation (40) as representing a pulsating mass (extended object of variable mass) with mean value $(\hbar/c) \|\mu\|$. If $\mathbf{B}$ is very large, we see that the effective mass can also be large. However, the operator still looks (almost) like the identity outside a few Compton wavelengths.

## VI. General Solution

In this section, we construct the solution of equation (1b) for the constant **A** case. The solution for the other examples will be discussed in a future paper on the application of these results to the canonical proper time formulation of relativistic quantum mechanics. First, rewrite equation (7) as:

$$Q(\mathbf{x},\mathbf{y};t) = \left(\frac{1}{4\pi t}\right)^{3/2} \exp\left\{-\frac{(\mathbf{x}-\mathbf{y})^2}{4t} - \mu^2 t + \frac{ie}{\hbar c}(\mathbf{x}-\mathbf{y})\cdot\mathbf{A}\right\}. \tag{42}$$

Now, using the theory of fractional powers, we note that if $\mathbf{T}[t,0]$ is the semigroup associated with $-\mathbf{G}+\omega^2$, then the semigroup associated with $\sqrt{-\mathbf{G}+\omega^2}$ is given by: (Appendix, equations A7, A10)

$$\mathbf{T}_{1/2}[t,0]\varphi(\mathbf{x}) =$$

$$\int_0^\infty \left\{\int_{\mathbf{R}^3} \left(\tfrac{1}{4\pi s}\right)^{3/2} \exp\left\{\left(-\tfrac{\|\mathbf{x}-\mathbf{y}\|^2}{4s} - \mu^2 s\right) + \frac{ie}{2\hbar c}(\mathbf{x}-\mathbf{y})\cdot\mathbf{A}\right\}\varphi(\mathbf{y})d\mathbf{y}\right\}\left\{\left(\tfrac{ct}{\sqrt{4\pi}}\right)\tfrac{1}{s^{3/2}}\exp\left(-\tfrac{(ct)^2}{4s}\right)\right\}ds. \tag{43}$$

From a table of Laplace transforms, we get that:

$$\int_0^\infty \exp\left(-\tfrac{a}{s} - ps\right)\frac{ds}{s^3} = 2\left(\frac{p}{a}\right)K_2\left[2(ap)^{1/2}\right]. \tag{44}$$

With $a = \left[\|\mathbf{x}-\mathbf{y}\|^2 + c^2 t^2\right]/4$, $p = \mu^2$, we can interchange the order of integration to get that

$$\mathbf{T}_{1/2}[t,0]\varphi(\mathbf{x}) =$$

$$\frac{ct}{4\pi^2}\int_{\mathbf{R}^3} \exp\left\{\frac{ie}{2\hbar c}(\mathbf{x}-\mathbf{y})\cdot\mathbf{A}\right\}\frac{2\mu^2 K_2\left[\mu\left(\|\mathbf{x}-\mathbf{y}\|^2 + c^2 t^2\right)^{1/2}\right]}{\left[\|\mathbf{x}-\mathbf{y}\|^2 + c^2 t^2\right]}\varphi(\mathbf{y})d\mathbf{y}. \tag{45}$$

We now use the fact that $\mathbf{T}_{1/2}[t,0]$ has a holomorphic extension into the complex plane so that we may compute the limit as $t \to it$. Setting $\mathbf{U}[t,0] = \lim_{\varepsilon\to 0}\boldsymbol{\beta}\mathbf{T}_{1/2}[(i+\varepsilon)t,0]$, we define:
(Note that $\exp\{\boldsymbol{\beta}it\sqrt{-\mathbf{G}+\omega^2}\} = \boldsymbol{\beta}\exp\{it\sqrt{-\mathbf{G}+\omega^2}\}$.)

$$\mathbf{Z}\left[\mu\left(c^2 t^2 - \|\mathbf{x}-\mathbf{y}\|^2\right)^{1/2}\right] = \frac{c t \boldsymbol{\beta}}{4\pi} \begin{cases} \dfrac{-H_2^{(1)}\left[\mu\left(c^2 t^2 - \|\mathbf{x}-\mathbf{y}\|^2\right)^{1/2}\right]}{\left[c^2 t^2 - \|\mathbf{x}-\mathbf{y}\|^2\right]}, & ct < -\|\mathbf{x}\|, \\[1em] \dfrac{-2i K_2\left[\mu\left(\|\mathbf{x}-\mathbf{y}\|^2 - c^2 t^2\right)^{1/2}\right]}{\pi\left[\|\mathbf{x}-\mathbf{y}\|^2 - c^2 t^2\right]}, & c|t| < \|\mathbf{x}\|, \\[1em] \dfrac{H_2^{(2)}\left[\mu\left(c^2 t^2 - \|\mathbf{x}-\mathbf{y}\|^2\right)^{1/2}\right]}{\left[c^2 t^2 - \|\mathbf{x}-\mathbf{y}\|^2\right]}, & ct > \|\mathbf{x}\|. \end{cases} \quad (46a)$$

Where, $H_2^{(1)}, H_2^{(2)}$ are the Hankel functions (see Gradshteyn and Ryzhik[12]). Then it follows that

$$\mathbf{U}[t,0]\varphi(\mathbf{x}) = \int_{\mathbf{R}^3} \mu^2 \exp\left\{\frac{ie}{2\hbar c}(\mathbf{x}-\mathbf{y})\cdot\mathbf{A}\right\} \mathbf{Z}\left[\mu\left(c^2 t^2 - \|\mathbf{x}-\mathbf{y}\|^2\right)^{1/2}\right]\varphi(\mathbf{y}) d\mathbf{y} \quad (46b)$$

solves

$$i\hbar \partial \psi(\mathbf{x},t)/\partial t = \left\{\boldsymbol{\beta}\sqrt{c^2\left(\mathbf{p}-\tfrac{e}{c}\mathbf{A}\right)^2 + m^2 c^4}\right\}\psi(\mathbf{x},t), \quad \psi(\mathbf{x},0) = \varphi(\mathbf{x}). \quad (47)$$

**VII. An Alternative Dirac, Square-root Connection**

Before Minkowski's postulate and his geometric interpretation of the special theory, it was not uncommon to associate the potential energy with the mass of the system. Since that time, it has been assumed that the potential energy should always be treated as the fourth component of a four-vector. This postulate has only been fruitful in the one-particle case (for both classical and quantum theory). It was first noted by Pryce[16] that, in the many-particle case, the canonical center-of-mass vector is not the fourth component of a four-vector so that a geometric interpretation is highly problematic. This is the major reason that there is not a (satisfactory) relativistic classical or quantum many-particle theory. A detailed review of the problems may be found in Gill, Zachary and Lindesay[17]. At the quantum level, Feynman's path integral formulation of quantum mechanics and his time-ordered operator formulation of quantum electrodynamics challenges Minkowski's geometric interpretation of the role for time. For a rigorous development of both these approaches, see Gill and Zachary[18,19].

Returning to equation (1), when the mass is constant it was shown by Case[20] that a Foldy-Wouthuysen[21] transformation $\left(U_{FW}^{-1}\mathbf{H}_s U_{FW} = \mathbf{H}_D\right)$ may be constructed for particles of spin 0 and 1, then Pursey[22] showed that a transformation $D$ exists for particles of arbitrary spin to map (1) into: (we focus on the spin 1/2 case)

$$D[\Psi] = \mathbf{H}_D \Psi = \left\{ c\boldsymbol{\alpha} \cdot \left(\mathbf{p} - \tfrac{e}{c}\mathbf{A}\right) + mc^2 \boldsymbol{\beta} \right\} \Psi. \tag{48}$$

In this section we show that, if we give up the Minkowski postulate, there is another relationship between the square-root and Dirac equations that also has a certain natural appeal.

First, we note that minimal coupling may also be introduced into the interacting Dirac operator via: [$mc^2\beta + V = \beta(mc^2 + \beta V)$, $\boldsymbol{\pi} = (\mathbf{p} - \tfrac{e}{c}\mathbf{A})$]

$$D[\Psi] = \left\{ c\boldsymbol{\alpha} \cdot \boldsymbol{\pi} + \beta(mc^2 + \beta V) \right\} \Psi. \tag{49}$$

If we treat the potential energy as a part of the mass term, then we can write equation (49) in the form (see Schiff[23], page 329):

$$\left\{ E + c\boldsymbol{\alpha} \cdot \boldsymbol{\pi} + \beta(mc^2 + \beta V) \right\} \Psi = \mathbf{0}. \tag{50}$$

Now, multiply on the left by $E - c\boldsymbol{\alpha} \cdot \boldsymbol{\pi} - \beta(mc^2 + \beta V)$, do the standard computations using $e\boldsymbol{\alpha} \cdot [E\mathbf{A} - \mathbf{A}E] = ie\hbar\boldsymbol{\alpha} \cdot [\partial \mathbf{A}/c\partial t]$, $c\beta\boldsymbol{\alpha} \cdot [\beta V \mathbf{p} - \mathbf{p}\beta V] = ie\hbar c \boldsymbol{\alpha} \cdot \nabla \varphi$ and get:

$$\left\{ E^2 - c^2\pi^2 + e\hbar c \boldsymbol{\Sigma} \cdot \mathbf{B} + ie\hbar c \boldsymbol{\alpha} \cdot \mathbf{E} + i\hbar\frac{\partial V}{\partial t} + 2i\hbar c\boldsymbol{\alpha} \cdot \nabla V - 2cV\boldsymbol{\alpha} \cdot \mathbf{p} - (mc^2 + \beta V)^2 \right\} \Psi = \mathbf{0}, \tag{51}$$

where $V = e\varphi$, and $\mathbf{E} = -\partial \mathbf{A}/c\partial t - \nabla \varphi$ is the electric field. The $2i e\hbar c\boldsymbol{\alpha} \cdot \nabla \varphi$ term occurs because we made $\varphi$ a part of the mass, which led to a sign change. Thus, this approach does not quite lead to an electric dipole moment as in the standard method. The Klein-Gordon (type) equation related to (50) (with the same eigenfunctions and eigenvalues) follows from (51):

$$-\hbar^2 \frac{\partial^2 \Psi}{\partial t^2} = \left\{ c^2\pi^2 + 2cV\boldsymbol{\alpha} \cdot \mathbf{p} - e\hbar c \boldsymbol{\Sigma} \cdot \mathbf{B} - ie\hbar c\boldsymbol{\alpha} \cdot \mathbf{E} - i\hbar\frac{\partial V}{\partial t} - 2i\hbar c\boldsymbol{\alpha} \cdot \nabla V + (mc^2 + \beta V)^2 \right\} \Psi. \tag{52}$$

For comparison, from Schiff[23] (equations (42.9) pg. 320 and (43.25), pg. 329) we have ($V = e\varphi$, $\partial \varphi/\partial t = 0$)

$$-\hbar^2 \frac{\partial^2 \Psi}{\partial t^2} - 2i\hbar c V \frac{\partial \Psi}{\partial t} = \left\{ c^2\pi^2 - e\hbar c \boldsymbol{\Sigma} \cdot \mathbf{B} - ie\hbar c\boldsymbol{\alpha} \cdot \mathbf{E} + i\hbar\frac{\partial V}{\partial t} + m^2 c^4 - V^2 \right\} \Psi.$$

Returning to (51), it is easy to see that:

$$E\Psi = \left\{ \beta \sqrt{c^2\pi^2 + 2cV\alpha \cdot \mathbf{p} - e\hbar c\Sigma \cdot \mathbf{B} - ie\hbar c\alpha \cdot \mathbf{E} - i\hbar \frac{\partial V}{\partial t} - 2i\hbar c\alpha \cdot \nabla V + (mc^2 + \beta V)^2} \right\} \Psi. \quad (53)$$

Note that equation (53) is not a diagonalized representation. However, it is an exact representation, which retains the same eigenfunctions and eigenvalues as the Dirac equation. Thus, we have another analytic representation for the Dirac equation (not a Foldy-Wouthuysen transformation[44]).

When **A** is time-independent and $V = 0$, we obtain a (diagonalized) form similar to one derived in deVries[24] via the Foldy-Wouthuysen transformation. In this situation, we obtain the most general (exact) representation that includes all time-independent vector potentials. (It is clear that, as expected, we can never obtain a diagonalized representation for nonzero $V$.)

While completing the final draft of this paper, we discovered a very interesting publication by Silenko[43] that includes a fairly complete review of the Foldy-Wouthuysen and other transformation methods, which seek to obtain consistent approximations to the Dirac equation in the relativistic and nonrelativistic cases.

In order to see how equation (53) relates to the Schrödinger equation, rewrite it as:

$$\left\{ \beta mc^2 \sqrt{1 + \beta \frac{2V}{mc^2} + \frac{V^2}{m^2c^4} + \frac{c\pi^2 + 2V\alpha \cdot \mathbf{p}}{m^2c^3} - \frac{e\hbar \Sigma \cdot \mathbf{B}}{m^2c^3} - \frac{ie\hbar \alpha \cdot \mathbf{E}}{m^2c^3} - \frac{i\hbar}{m^2c^4} \frac{\partial V}{\partial t} - \frac{2i\hbar \alpha \cdot \nabla V}{m^2c^4}} \right\} \Psi.$$

Expand the above equation to first order to get:

$$S[\Psi] \cong \left\{ \beta mc^2 + V + \frac{\beta V^2}{2mc^2} + \frac{\beta \pi^2}{2m} + \frac{V\alpha \cdot \mathbf{p}}{mc^2} - \frac{e\hbar \beta \Sigma \cdot \mathbf{B}}{2mc} - \frac{ie\hbar \beta \alpha \cdot \mathbf{E}}{2mc} - \frac{i\hbar \beta \partial V}{mc^2 \partial t} - \frac{i\hbar}{mc} \beta \alpha \cdot \nabla V \right\} \Psi. \quad (54)$$

It follows that, dropping terms of order $(v/c)^2$, the corresponding (positive energy) Schrödinger equation is (see Schiff[23], page 330, last paragraph)

$$i\hbar \frac{\partial \Psi}{\partial t} = \left\{ \frac{\pi^2}{2m} + V + mc^2 - \frac{e\hbar}{2mc} \Sigma \cdot \mathbf{B} + V^2/2mc^2 \right\} \Psi. \quad (55)$$

**Perturbations**

It is clear that, in general, many problems of interest will have time-dependent vector potentials. In this section, we develop the basics of a perturbation theory for the square root operator. To begin, we assume that the vector potential can be decomposed as $\mathbf{A} = \mathbf{A}_1 + \mathbf{A}_2$, where $\mathbf{A}_1$ is the potential for a constant magnetic field, while $\mathbf{A}_2$ may have arbitrary space and time dependence. We first write $\left[\mathbf{p} - (e/c)\mathbf{A}\right]^2$ as

$$\left(\mathbf{p} - \tfrac{e}{c}\mathbf{A}_1\right)^2 - \left(\mathbf{p} - \tfrac{e}{c}\mathbf{A}_1\right)\cdot \mathbf{A}_2 - \mathbf{A}_2 \cdot \left(\mathbf{p} - \tfrac{e}{c}\mathbf{A}_1\right) + |\mathbf{A}_2|^2.$$

Using this result, write equation (53) as $\beta\sqrt{c^2\left[\mathbf{p} - (e/c)\mathbf{A}_1\right]^2 + m^2c^4 - e\hbar c \Sigma \cdot \mathbf{B}_1 + F}$, where

$$F = \left(mc^2 + \beta V\right)^2 - m^2c^4 + |\mathbf{A}_2|^2 - \left(\mathbf{p} - \tfrac{e}{c}\mathbf{A}_1\right)\cdot \mathbf{A}_2 + 2cV\alpha \cdot \mathbf{p}$$
$$- \mathbf{A}_2 \cdot \left(\mathbf{p} - \tfrac{e}{c}\mathbf{A}_1\right) - e\hbar c \Sigma \cdot \mathbf{B}_2 - ie\hbar c \alpha \cdot \mathbf{E} - i\hbar \frac{\partial V}{\partial t} - 2i\hbar c \alpha \cdot \nabla V.$$

We now assume that the operator $G = c^2\left[\mathbf{p} - (e/c)\mathbf{A}_1\right]^2 + m^2c^4 - e\hbar c \Sigma \cdot \mathbf{B}_1$ has the property that $G^{-1/2}F$ is bounded. When this condition is satisfied, we can expand (53) as:

$$\begin{aligned}\beta\sqrt{G+F} &= \beta\sqrt{G}\sqrt{I + G^{-1}F} = \beta\sqrt{G}\sum_{n=0}^{\infty}\binom{1/2}{n}\left(G^{-1}F\right)^n \\ &= \beta\left\{\sqrt{G} + \tfrac{1}{2}G^{-1/2}F - \tfrac{1}{8}G^{-1/2}FG^{-1}F + \cdots\right\}\end{aligned} \quad (56)$$

The assumption insures that equation (56) converges rapidly so that the first few terms give a good approximation. Note that all the terms are computable so that equation (56) has more than theoretical value.

**Conclusion**

In this paper, we have shown that the square-root operator has a well-defined analytic representation, which is uniquely determined by the Green's function for the corresponding Schrödinger equation for all values of the spin (our focus was on spin 1/2). We have constructed the exact solution and have explored a number of simple cases in order to obtain some insight into the physical meaning of the operator. In the free case, the operator has a representation as a nonlocal composite of three singularities. To our knowledge, this is the first example of a physically relevant operator with these properties. In the standard interpretation, the particle component has two negative parts and one (hard core) positive part, while the antiparticle component has two positive parts and one (hard core) negative part. This effect is confined within a Compton wavelength such that, at the point of singularity, they cancel each other providing a finite result. Furthermore, the operator looks (almost) like the identity outside a Compton wavelength, but has a residual instantaneous connection with all other particles in the universe at each point in time. We thus suggest that this operator represents the inside of an extended object.

In addition to the free particle, we have considered the case of a constant vector potential and a constant magnetic field. They both reveal the complex nature of the internal dynamics when interaction is turned on. When the vector potential is constant but nonzero, we get two extra terms and a multiplicative exponential factor. One term is real-valued, while the other is purely imaginary, and both are nonsingular.

The next case explored corresponds to a constant magnetic field. In this case, we get two extra terms but no multiplicative exponential factor. The real-valued term is the same as in the case of a constant vector potential. However, the purely imaginary term becomes highly singular and the effective mass becomes complex. The physical picture that emerges is that, in addition to the confined singularities and the additional attractive (repulsive) term, the effective mass of the composite acquires an oscillatory behavior.

We also constructed the complete propagator for the constant vector potential case. Then we studied an alternative relationship between the Dirac equation (with minimal coupling) and the square-root equation that is much closer than the conventional one. This was accomplished by considering the scalar potential as a part of the mass. This allowed us to derive a new Klein-Gordon equation and a new square-root equation that both have the same eigenvalues and eigenfunctions as the Dirac equation.

In the last section we developed a perturbation formula which allows us to consider suitable spacetime-dependent vector and scalar potentials. We have included an appendix with a brief review of the theory of semigroups of operators relevant for the construction of our analytic representation of the square-root operator.

There are a number of issues that we have not discussed in this paper. Many writers have used the square-root operator to develop constituent quark models. These models are very accurate in the description of a large part of meson and baryon properties (see Brau[25] and references therein). The work of Sucher[26] suggests that, with minimal coupling, the square-root operator may not be Lorentz invariant, while Smith[27] suggests that the equation has very limited gauge properties. These are very complicated problems that require additional study and analysis.

If our conclusion is correct that the square-root operator represents the inside of a particle, then the question of Lorentz invariance may be mute and all assumptions would require careful reconsideration. First, recall that the special theory of relativity has serious problems when applied to nonpoint particles (extended objects). Furthermore, a number of writers have suggested the possible breakdown of Lorentz invariance inside extended (nonlocal) objects. Indeed, sometime ago, Santilli[28-31] called for detailed experimental study to determine if the special theory was still valid inside a hadron and/or hot hyper-dense matter such as a star. Coleman and Glashow (see[32] and references therein) have identified possible terms within the perturbative framework of the standard model which would allow small departures from Lorentz invariance. Their interest is in identifying phenomena which could be relevant to both cosmic and neutrino physics. At the cosmic level their hope is to undo the GZK cutoff for high-energy cosmic rays; while at the neutrino level, the hope is to identify novel types of neutrino oscillations. The varying speed of light theory (VSL) of Moffat[33] provides an elegant solution to a number of cosmological problems: the horizon, flatness, and lambda problems of big-bang cosmology (see also Magueijo[34]). In closing, we note that the strongest experimental (and phenomenological) analysis suggesting deviations from Lorentz symmetry inside hadrons is that due to Arestov et. al[35].

Finally, in addition to the obvious possibility that the square-root operator may be used to represent the inside of hadrons, it is also possible that the residual attractive (particle) part may be the long-sought cause for the gravitational interaction in matter. If this view is correct, then we would expect matter and antimatter to be gravitationally attractive among themselves and gravitationally repulsive to each other. This would make physical sense if we take seriously the interpretation of antimatter as matter with its time reversed (as opposed to hole theory). Such a hypothesis is within current experimental capability. Indeed, Santilli[31] has also suggested this possibility and he has identified relevant test experiments. It should be noted that, in the quoted references, Santilli avoided the standard (historical) objections to antigravity by introducing a new class of numbers with a negative unit, and called them isoduals of the standard number system. Subject to a more detailed study, it appears that the antimatter component of our analysis can be easily reinterpreted via Santilli's isodual theory without any change in the results.

In this sense, our analysis constitutes a possible theoretical confirmation of the axiomatic prediction of antigravity for matter-antimatter systems as suggested above.

**Appendix: Semigroups and Fractional Powers of Operators**

This appendix provides a brief survey of the theory of strongly continuous semigroups of linear operators, which is used to explain the general theory of fractional powers of operators. The definitions and basic results are recorded here for reference so as to make the paper self-contained. Hille and Phillips[36], and Yosida[37] are the general references on semigroups (see also Goldstein[38], Engel and Nagel[39] and Pazy[40]). Butzer and Berens[41] have a very nice (short) introduction to operator semigroups. Tanabe[42] has a good section on fractional powers, but one should also consult Yosida[37].

**Definition A.1** Let $\mathbf{T}(t)$, $t \geq 0$ be a family of bounded linear operators on a Banach space $\mathbf{B}$. This family is called a strongly continuous semigroup of operators (or a $C_0$-semigroup) if the following conditions are satisfied:

1. $\mathbf{T}(t+s) = \mathbf{T}(t)\mathbf{T}(s) = \mathbf{T}(s)\mathbf{T}(t)$, $\forall t, s \geq 0$, $\mathbf{T}(0) = \mathbf{I}$,

2. $\lim_{t \to s} \mathbf{T}(t)\varphi = \mathbf{T}(s)\varphi$, $\forall \varphi \in \mathbf{B}$.

If the family $\mathbf{T}(t)$ is defined for $t \in (-\infty, \infty)$, then it is called a $C_0$-group and $\mathbf{T}(-t) = \mathbf{T}^{-1}(t)$. By further restriction, we obtain the (well-known) definition of a unitary group.

**Theorem A.2** Let $\mathbf{T}(t)$, $t \geq 0$, be a $C_0$-semigroup and let $D = \{\varphi \mid \lim_{h \to 0} h^{-1}[\mathbf{T}(h) - \mathbf{I}]\varphi \text{ exists}\}$. Define $A\varphi = \lim_{h \to 0} h^{-1}[\mathbf{T}(h) - \mathbf{I}]\varphi$ for $\varphi \in D$; then:

1. $D$ is dense in $\mathbf{B}$,

2. $\varphi \in D \Rightarrow \mathbf{T}(t)\varphi \in D$, $t \geq 0$,

3. $\dfrac{d}{dt}[\mathbf{T}(t)\varphi] = A\mathbf{T}(t)\varphi = \mathbf{T}(t)A\varphi$, $\forall \varphi \in D.$ (A1)

**Proof:** (see Tanabe[42], page 51)

Thus, it follows that $\psi(\mathbf{x}, t) = \mathbf{T}(t)\varphi(\mathbf{x})$ solves the initial value problem:

$$\frac{d}{dt}\psi(\mathbf{x}, t) = A\psi(\mathbf{x}, t), \psi(\mathbf{x}, 0) = \varphi(\mathbf{x}).$$ (A2)

The operator $A$ is called the generator of the semigroup $\mathbf{T}(t)$, and we can write

$$\mathbf{T}(t)\varphi(\mathbf{x}) = \exp\{tA\}\varphi(\mathbf{x})$$

**Theorem A.3** The generator $A$, of the semigroup $\{\mathbf{T}(t), t \geq 0\}$ is a closed linear operator. If $\|\mathbf{T}(t)\| \leq M\exp\{\beta t\}$, for fixed constants $M$ and $\beta$, then the half-plane $\{\lambda \mid \operatorname{Re}(\lambda) > \beta\}$ is contained in the resolvent set $\rho(A)$ and, for each such $\lambda$, we have: (Tanabe[42], page 55)

$$(\lambda \mathbf{I} - A)^{-1}\varphi = \int_0^\infty e^{-\lambda t}\mathbf{T}(t)\varphi\, dt = \mathbf{R}(\lambda, A). \tag{A3}$$

$\mathbf{R}(\lambda, A)$ is called the resolvent operator of $A$ and

$$\|\mathbf{R}(\lambda, A)\| \leq M[\operatorname{Re}(\lambda) - \beta]^{-1}. \tag{A4}$$

**Definition A.4** Suppose that the operator $A$ generates a $C_0$-semigroup on $\mathbf{B}$, and there exists a constant $M$ such that: (see Engel and Nagel[39], page 101, Theorem 4.6)

$$\|\mathbf{R}(r + is, A)\| \leq \frac{M}{|s|}, \forall r > 0 \text{ and } 0 \neq s \in \mathbf{R}^1. \tag{A5}$$

Let $\Sigma$ represent the above region in the complex plane, then the family $\{\mathbf{T}(z), z \in \Sigma\}$ is a called a holomorphic $C_0$-semigroup on $\mathbf{B}$. (See Engel and Nagel[39] for details.)

Introduce the function $f_{t,\alpha}(\lambda)$ defined by: (Yosida[37], page 259)

$$f_{t,\alpha}(\lambda) = \frac{1}{2\pi i}\int_{\sigma-i\infty}^{\sigma+i\infty} \exp\{z\lambda - tz^\alpha\}\, dz,\ \lambda \geq 0,$$
$$f_{t,\alpha}(\lambda) = 0,\ \lambda < 0, \tag{A6}$$

where $t > 0$, $0 < \alpha < 1$ and $\sigma > 0$, and the branch of $z^\alpha$ is taken so that $\operatorname{Re}(z^\alpha) > 0$ when $\operatorname{Re}(z) > 0$. The branch is a single-valued function in the complex plane cut along the negative real axis. The convergence of the integral (A6) is insured by the factor $\exp\{-tz^\alpha\}$. Define $\mathbf{T}_\alpha(t)$ by $\mathbf{T}_\alpha(0)\varphi = \varphi$ and for $t > 0$,

$$\mathbf{T}_\alpha(t)\varphi = \int_0^\infty f_{t,\alpha}(s)\mathbf{T}(s)\varphi\, ds, \tag{A7}$$

where $\{\mathbf{T}(t), t \geq 0\}$ is a $C_0$-semigroup of operators on $\mathbf{B}$.

**Theorem A.5** Suppose that the operator $A$ generates a $C_0$-semigroup $\{\mathbf{T}(t), t \geq 0\}$ on $\mathbf{B}$. Then:
1. the family $\{\mathbf{T}_\alpha(t), t \geq 0\}$ is a holomorphic $C_0$-semigroup on $\mathbf{B}$,

2. the operator $A_\alpha$, the generator of $\{\mathbf{T}_\alpha(t),\ t \geq 0\}$, is defined by: $A_\alpha \varphi = -(-A)^\alpha \varphi$, and

$$A_\alpha \varphi = \frac{\sin \alpha \pi}{\pi} \int_0^\infty \lambda^{\alpha-1} R(\lambda, A)[-A\varphi] d\lambda. \tag{A8}$$

**Proof:** (see Yosida[37], page 260).

For our work, we are only interested in the case when $\alpha = 1/2$. Let us deform the path of integration in equation (A6) into a union of two paths, $re^{-i\theta}$, when $-r \in (-\infty, 0)$ and $re^{i\theta}$, when $r \in (0, \infty)$, where $\pi/2 \leq \theta \leq \pi$. In particular, we need $\theta = \pi$. This case leads to:

$$f_{t,1/2}(s) = \frac{1}{\pi} \int_0^\infty \exp\{-sr\} \sin\{tr^{1/2}\} dr. \tag{A9}$$

Using a table of Laplace transforms, we have

$$f_{t,1/2}(s) = \frac{ts^{-3/2}}{\sqrt{4\pi}} \exp\{-\frac{t^2}{4s}\}. \tag{A10}$$